\documentclass[10pt,a4paper]{article}

    \usepackage[T1]{fontenc}
    \usepackage{inputenc}
    \usepackage[english]{babel}
    \usepackage[nohead,includefoot,margin=2cm]{geometry}
    \usepackage[compact,raggedright,small]{titlesec}
    \usepackage[affil-it]{authblk}
    \usepackage[runin]{abstract}
    \usepackage{multicol}
    \usepackage{blindtext}
\usepackage{subfig}
\usepackage{graphicx}
\usepackage{cite}

    \newcommand*{\pacsname}{PACS numbers}

    \addto{\Affilfont}{\small}

    \title{\large\bfseries Spatial Integration by a Dielectric Slab Waveguide and its Planar Graphene-based Counterpart}

    \author[1]{Farzad Zangeneh-Nejad}
    \author[1,*]{Amin Khavasi}
    \affil[1]{Department of Electrical Engineering, Sharif University of Technology, Tehran, Iran, P.O. Box 11555-4363}
\affil[*]{Corresponding author: khavasi@sharif.edu}
    
\begin{document}
       \date{}
       \maketitle

       \begin{abstract}
Motivated by the recent progress in analog computing [Science 343, 160 (2014)], a new approach to perform spatial integration is presented using a dielectric slab waveguide. Our approach is indeed based on the fact that the transmission coefficient of a simple dielectric slab waveguide at its mode excitation angle matches the Green's function of first order integration. Inspired by the mentioned dielectric-based integrator, we further demonstrate its graphene-based counterpart. The latter is not only reconfigurable but also highly miniaturized in contrast to the previously reported designs [Opt. Commun. 338, 457 (2015)].  Such integrators have the potential to be used in ultrafast analog computation and signal processing.
       \end{abstract}

\indent The idea of spatial computation fundamentally originates from traditional analog computers, or "calculating machines", which were made to perform mathematical operations electronically or mechanically. Such machines were actually limited by their relatively large size and slow response \cite{1,2}. However, very recently, researchers have succeeded to overcome theses restrictions by exploiting the recent advances in optical technology and by  introducing a new concept, namely optical spatial computing \cite{3,4,5,6,7,8,9,10}.\\
\indent The ideas to perform optical spatial computing can be split into two groups. In the first approach, namely metasurface (MS) approach, the computation is performed using a properly designed metasurface filter performing the Green's function of the operator of choice in the spatial Fourier domain \cite{4}. Despite the applicability of this approach, the necessity of two additional sub-blocks to apply Fourier and Inverse Fourier Transform before and after the main block, i.e.  metasurface filter, causes the the overall size of such systems  to be high. This reduces the practical worth of this method \cite{3}.\\
\indent In another approach, namely Green's Function (GF) method,  the spatial computing is performed by properly adjusting the thicknesses and refractive indexes of individual layers of a multilayered slab which is homogenous along two directions (for example along x and z axes). The mathematical operation is directly realized through processing the optical field as it travels through each layer along the other direction (i.e. y axis). By this technique, one avoids going into the Fourier domain, and consequently avoids any further fabrication complexity \cite{3}. However, this approach has two major weaknesses too. For one thing, due to the reflection symmetry of such multilayered slab, it is not possible to realize many important operators whose green's functions have an odd symmetry in the spatial Fourier domain, like the operator of first order differentiation or integration.  Moreover, in this technique, the associated values of refractive indexes and thicknesses of slabs are calculated making use of an optimization method which not always results in practical values \cite{3,6}.\\
\indent Very recently, by breaking the reflection symmetry of the aforementioned multilayered slab, and by employing Brewster effect, we demonstrated a practical first order differentiator for the first time \cite{6}. Our proposed Brewster differentiator worked based on the fact that near the zero reflection for obliquely incident waves at Brewster angle, the spatial impulse response of the structure could be approximated with a linear function which matches the Green's function of the spatial differentiation well.\\
\indent In this contribution, our aim is to propose another novel and practical idea, this time to demonstrate a first order integrator. To this end, we first show that the transmission coefficient of a dielectric slab waveguide has a pole at the incident angle for which the mode of the waveguide is excited. We then base our integrator on the fact that near the mentioned pole, the spatial impulse response of such structure can be approximated with a function which matches the Green's function of the spatial integration well. Motivated by the great performance of the latter integrator, and by exploiting the unique features of promising graphene regarding its tunable conductivity and ability to confine modes in highly small volumes, we then demonstrate a tunable, miniaturized and planar spatial integrator. Such integrators have the potential to be used in ultrafast analog computation, equation solving, and signal processing.\\
\indent We first realize the spatial integrator using a dielectric slab waveguide. To this end, the primary sketch of our proposed approach is provided in Fig. 1 (a). As it is observed, an input field with the profile  $f(y)$ is considered to be obliquely incident onto the boundary of a dielectric layer with refractive index of $n_{2}$, which in turn, is sandwiched between two semi-infinite dielectric media with refractive indexes of $n_{1}$. Based on the fact that the Green's function of a first-order integrator is of the form $G(k_{y})=1/ik_{y}$, the transmission or reflection coefficient of the structure of Fig. 1 (a) has to have a first order pole at at least one incident angle so that such a system can perform integration at that angle.\\
 \indent It is well-known that at the incident angle for which the mode of the structure of Fig. 1 (a) is excited, the input field is coupled to the slab, and the transmission coefficient of the structure will have a first-order pole thereafter \cite{11}. As described in \cite{6}, if the corresponding transmission coefficient is considered as the Green's function, the output of such system, i.e. the transmitted field, will then be first order integration of the input field.\\
 \begin{figure}
\centering
     \subfloat[\label{subfig-1:dummy}]{%
       \includegraphics[width=0.35\textwidth]{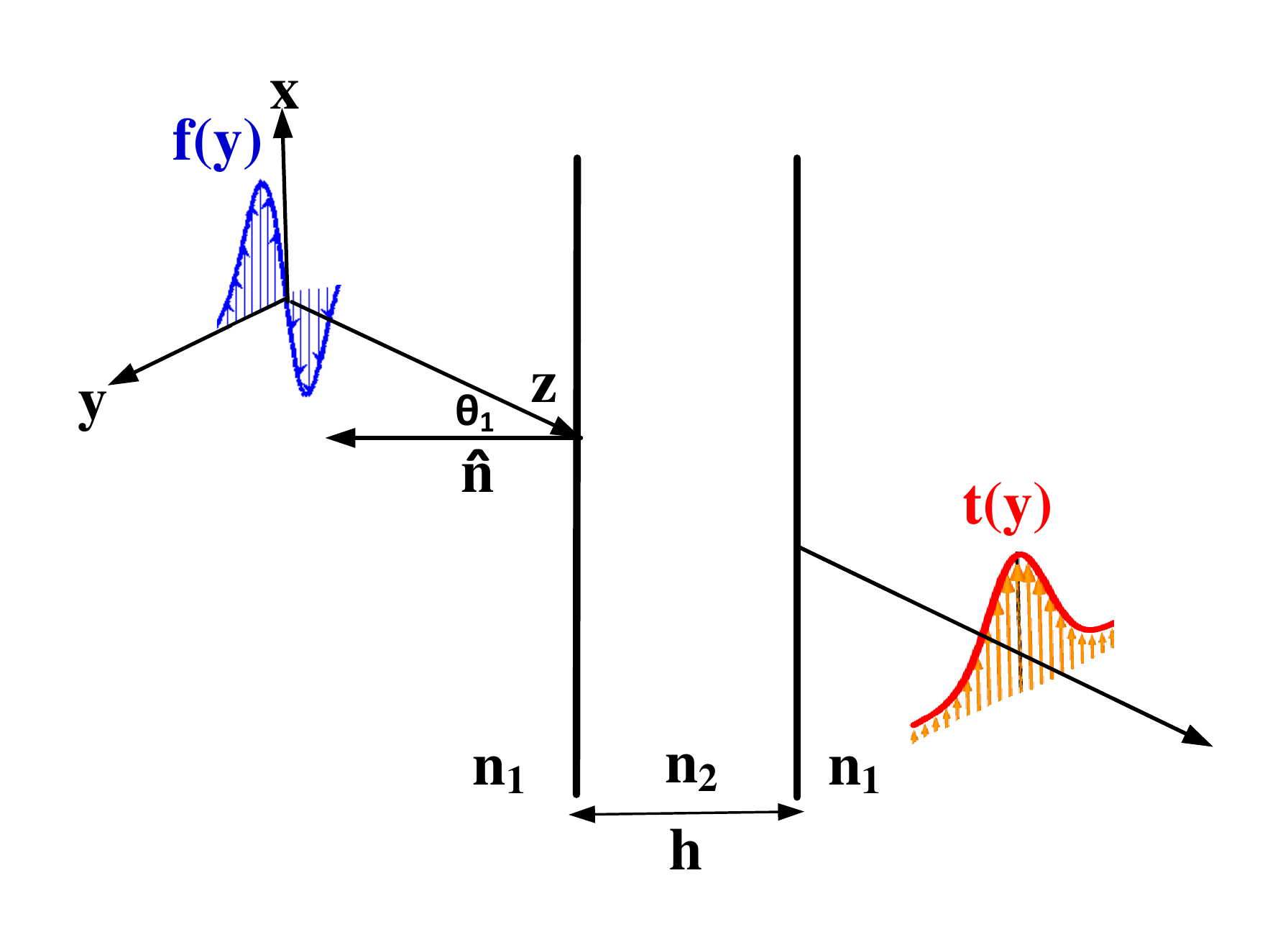}
     }

     \subfloat[\label{subfig-2:dummy}]{%
       \includegraphics[width=0.37\textwidth]{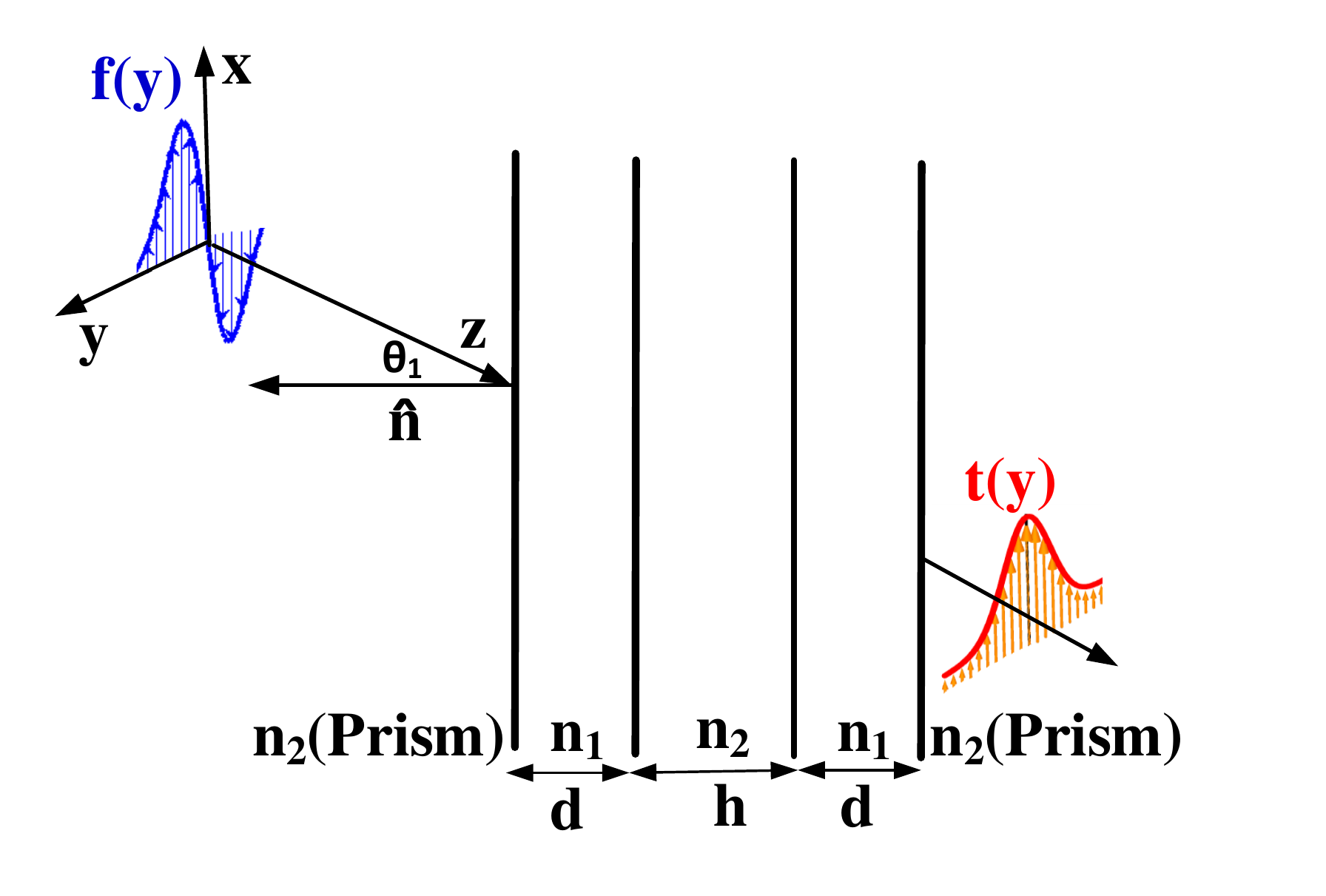}
     }
     \caption{(a) Primary structure of the proposed approach to perform integration using mode excitation of a dielectric slab waveguide (b) Modified structure of the integrator: Prism coupling technique \cite{12} has been used to excite the mode of the waveguide}
     \label{fig:dummy}
\end{figure}
 \indent Although the proposed structure in Fig. 1 (a) works well in theory, the situation in practice is rather different. In fact, in order to have a guided mode in such an integrator,  the parameter $n_{2}$ has to be higher than $n_{1}$, i.e. $n_2>n_1$. In this case, the phase matching criteria directly implies that the angle of incidence must have a non-zero imaginary part so that such a mode can be excited, which is not possible in reality \cite{12}. Because of this weakness, we develop our primary integrator shown in Fig. 1 (a) to a modified one in Fig. 1 (b). In the modified integrator, a well-known phase-matching technique known as prism coupling \cite{12} has been used to excite the mode of the structure. As it can been seen from Fig. 1 (b), two additional semi-infinite dielectric layers with the refractive indexes of $n_{2}$ are added to the beginning and the end of the primary integrator so as to be able to excite the mode of the waveguide with a real value of incident angle. It should be noted that the modes excited via this technique are inherently leaky which means that the they lose energy due to the leakage of radiation into the prism, i.e. to the two added dielectric layers. We will show that this makes the transmission coefficient of the modified integrator be limited to one (instead of infinity) at the angle of mode excitation. In addition, this phenomena which is consistent with the passivity of the proposed device, makes the proposed integrator be not able to retrieve the zero spatial harmonic of the input field. \\
 \indent As a specific example of the proposed integrator of Fig. 1 (b), we assume $n_{1}=1.5$ (the refractive index of SiO$_{2}$ at optical range), $n_{2}=3.4$ (the refractive index of Si at optical range), $h=0.4~\mu m$ and $d=0.2~ \mu m$. Moreover, we consider the operation wavelength of the integrator to be $\lambda_{0}=1.5~\mu m$. To have a mode in such a structure the following criteria must be maintained \cite{11}
 \begin{eqnarray}\label{1}
\theta_{1}>\theta_{c}=sin^{-1}(\frac{n_{1}}{n_{2}})\\ \nonumber
\psi=\frac{2\pi}{\lambda_{0}}n_{2}L_1cos(\theta_{1})-\varphi_{c}=\upsilon \pi
 \end{eqnarray}
 in which $\upsilon$ is an integer and $\varphi_{c}$ is of the form
 \begin{equation}\label{2}
 \varphi_{c}=tan^{-1}(F\sqrt{\frac{(n_1sin(\theta_1))^{2}-n_2^{2}}{n_1cos(\theta_1)}})
 \end{equation}
   where $F=1$ for TE and $F=n_1^2/n_2^2$ for TM polarized incident waves. By solving Eqs. 1 and 2, one obtains that the criteria of Eq. 1 is established at mode excitation angles of $\theta_{TE}=66.9^{\circ}$ for TE, and $\theta_{TM}=59.7^{\circ}$ for TM case.\\
  \indent The transmission coefficient of the structure and its associated Green's function $G(k_y)$ for the TE case are then calculated and depicted in Figs. 2 (a) and (b), respectively. It is obvious that the corresponding transmission coefficient has a first order pole at the angle $\theta_{TE}$, as the behaviour of its related Green's function is just like that of an ideal integrator at this angle. However, due to the previously explained reason, its associated value is limited to 1. As we will see in the following, this prevents the integrator from retrieving the zero spatial harmonic of the input field properly. The corresponding transmission coefficient and Green's function $G(k_y)$ for the TM case are also indicated in Figs. 2 (c) and (d). It can be seen that Green's functions $G(k_y)$ can be well-estimated by the Green's function of an ideal integrator for both major polarizations, provided that the spatial bandwidth of the input field is small.
\begin{figure}
\centering
     \subfloat[\label{subfig-1:dummy}]{%
       \includegraphics[width=0.24\textwidth]{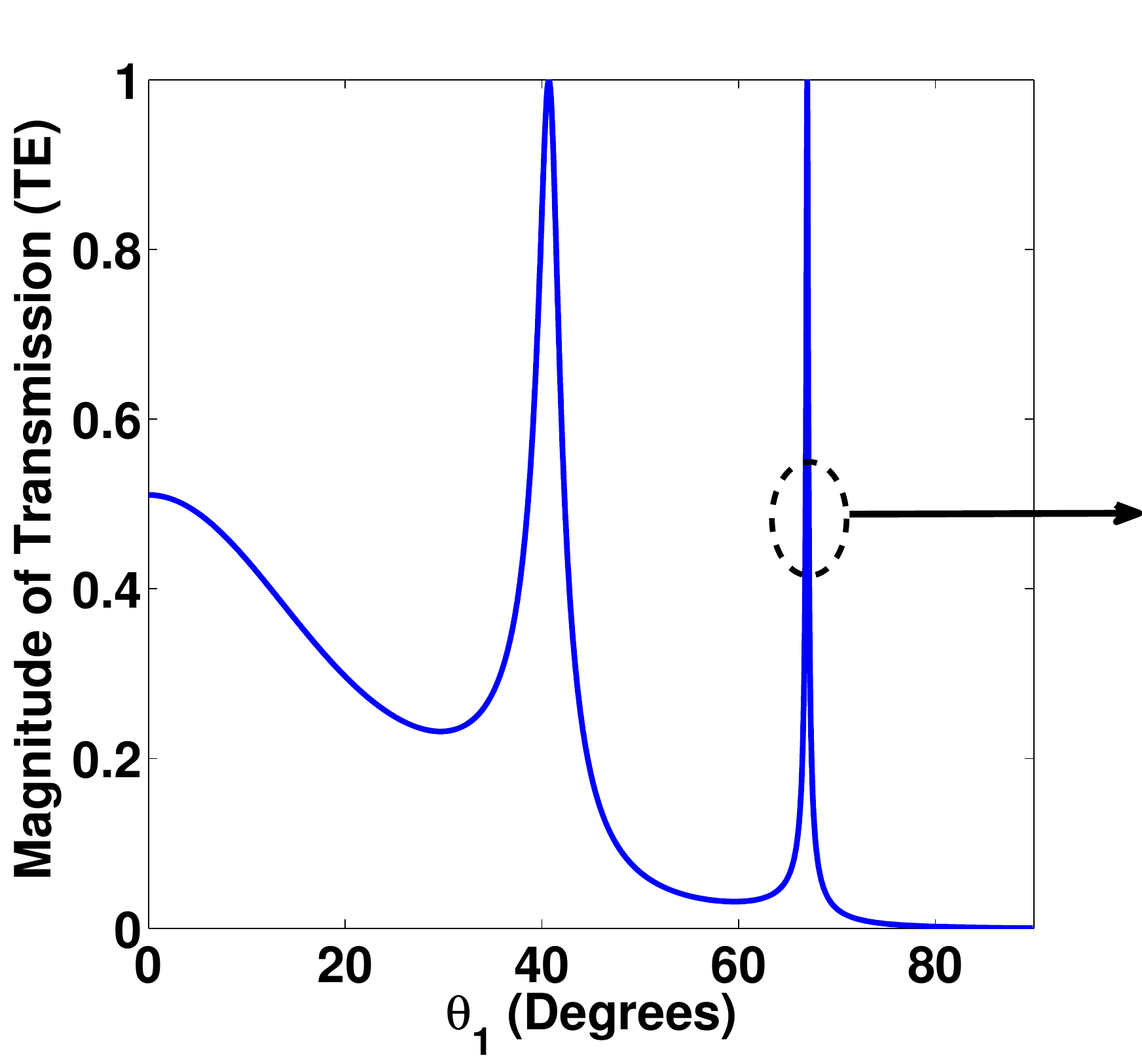}
     }
          \subfloat[\label{subfig-1:dummy}]{%
       \includegraphics[width=0.24\textwidth]{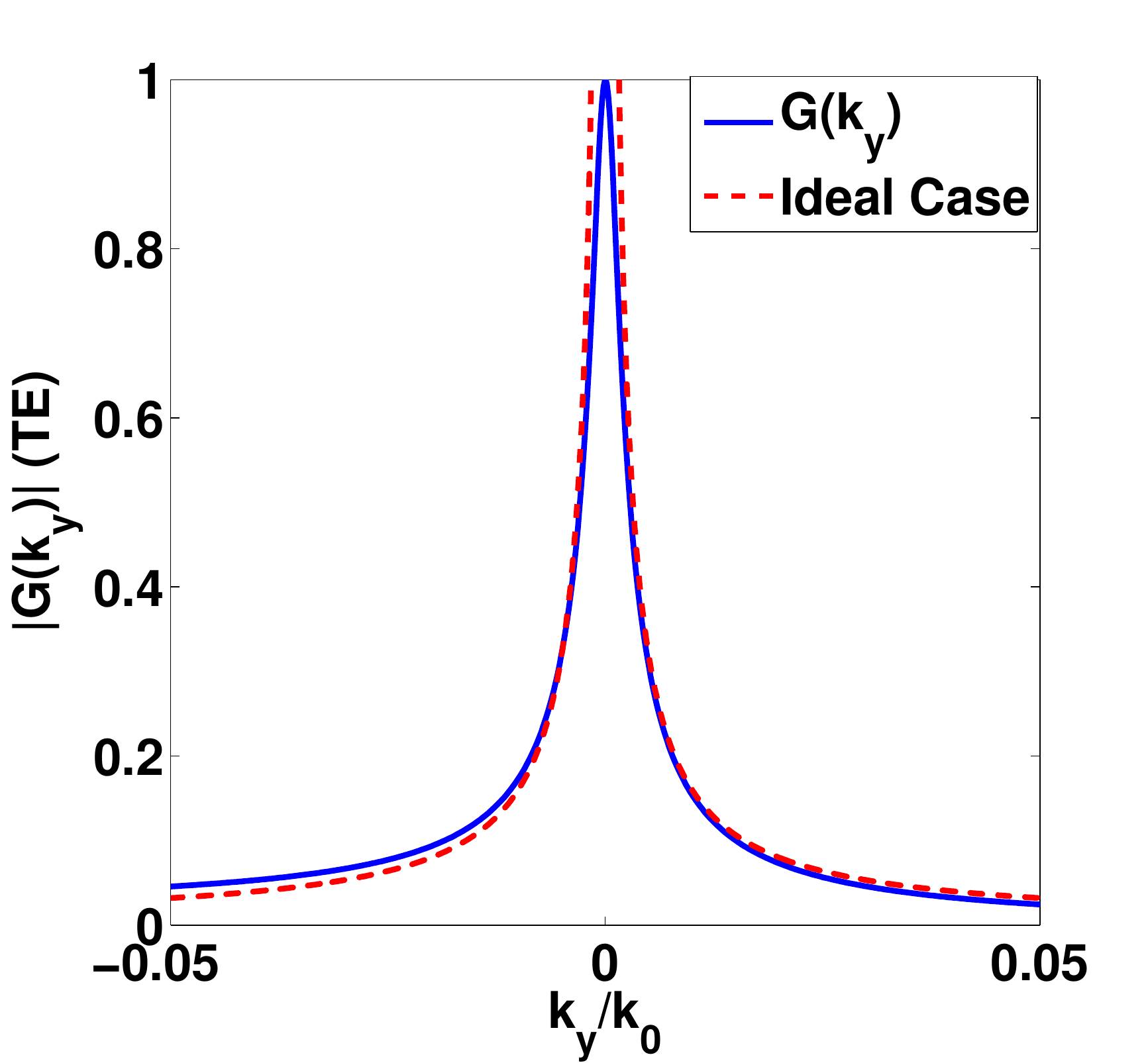}
     }
     \hfill
     \subfloat[\label{subfig-2:dummy}]{%
       \includegraphics[width=0.24\textwidth]{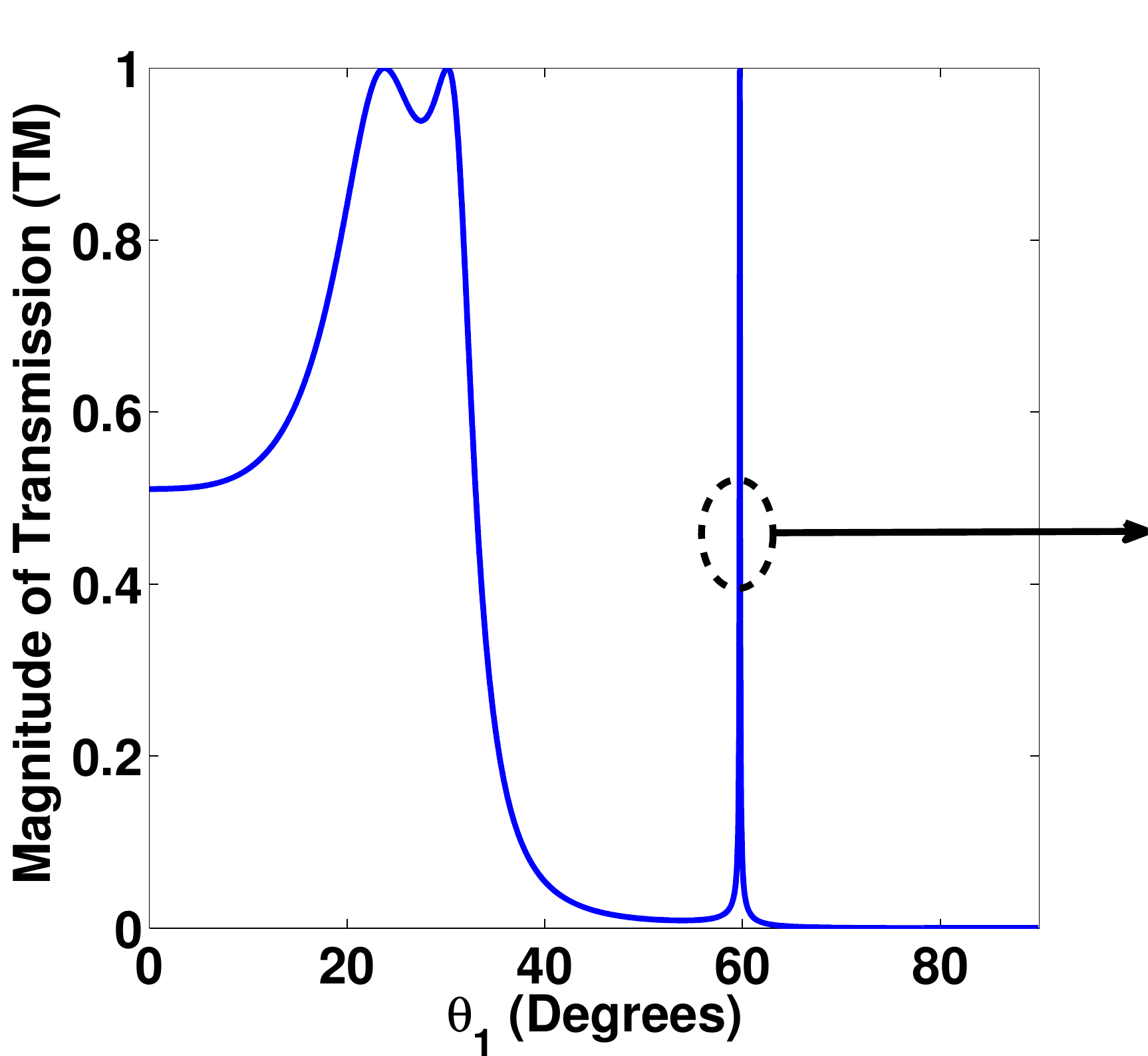}
     }
     \subfloat[\label{subfig-2:dummy}]{%
       \includegraphics[width=0.24\textwidth]{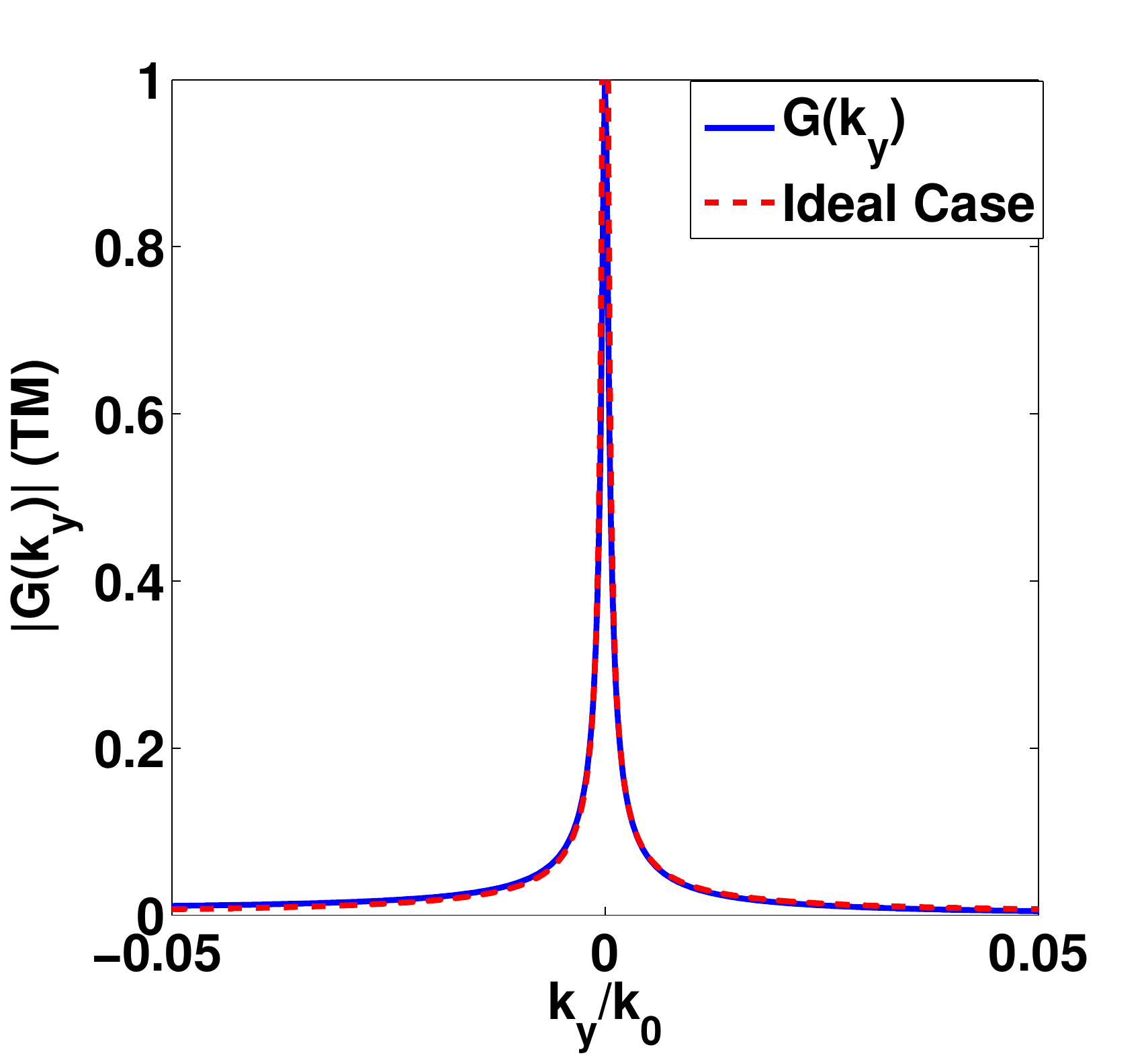}
     }
     \caption{Transmission Coefficients and Green's functions of the dielectric integrator compared with the Green's function of an ideal integrator for TE case ((a), (b)) and TM case ((c),(d))}
     \label{fig:dummy}
\end{figure}

\indent To evaluate the presented integrator, we first consider a TE polarized Gaussian incident field with a spatial bandwidth of $W =0.1k_{0}$ to be incident on the structure at the angle of $\theta_{TE}$. Fig. 3 (a) indicates the transmitted field $t(y)$ compared with the scaled exact first order integration of the input filed. As it is observed, while the integrator has worked well to retrieve the non-zero spatial harmonics of the input field, it fails to retrieve the zero harmonic or the DC part. As another example, we assume an input field with a Sinc profile and a spatial bandwidth of $W =0.1k_{0}$ whose its zero harmonic is eliminated. Fig. 3 (b) shows the transmitted field together with the exact integration of the input filed. In this case, since the input field has no zero spatial harmonic, the results match each other much better than the previous case.  The results for TM polarized incident fields with similar  profiles are also calculated and shown in Figs. 3 (c) and (d), respectively.\\
\indent Inspired by the proposed dielectric integrator of Fig. 1 (b), and by taking the advantage of tunability of graphene surface conductivity, we now turn to a reconfigurable, highly miniaturized, and planar graphene-based integrator. To this end, we only replace the dielectric layers with refractive indexes of $n_1$ and $n_2$ with 2D graphene sheets with surface conductivities of $\sigma_{g_{1}}$ and $\sigma_{g_{2}}$. We show that the latter integrator has no dependency to the angle of incidence as it can be adapted to any arbitrary incident angle simply  by tuning the surface conductivities of graphene films via a gate voltage \cite{14}. Furthermore, such an integrator is highly miniaturized due to the fact that it works based on graphene plasmons (GP) modes whose propagation constants are much higher than that of free excitation. The propagation constant of graphene plasmons $\beta_{p}$ can be obtained by a quasi-static approximation as \cite{13}
\begin{equation}\label{3}
 \beta_{p_{1,2}}=\frac{2i\omega\varepsilon_{0}}{\sigma_{g_{1,2}}}
\end{equation}
where $\omega$ is angular frequency and $\sigma_{g}$ is the graphene surface conductivity, which in turn can be calculated using Kubo formula \cite{14}.\\
\indent As an example of the proposed graphene-based integrator, we assume typical values of $\beta_{p_{1}}=1~\mu m^{-1}$ and $\beta_{p_{1}}=1.5~\mu m^{-1}$ for the propagation constants of graphene plasmons at the operational wavelength $\lambda_{0}=100 ~ \mu m$  \cite{13}. Moreover, we assume $h=d=3~\mu m$. Similar to the dielectric case, to have a plasmonic guided mode in such a structure the following criteria must be established
\begin{eqnarray}\label{1}
\theta_{1}>\theta_{c}=sin^{-1}(\frac{\sigma_{g_{2}}}{\sigma_{g_{1}}})=sin^{-1}(\frac{\beta_{p_{1}}}{\beta_{p_{2}}})\\ \nonumber
\psi_g=\beta_{p_{2}}Lcos(\theta_{1})-\varphi_{c}=\upsilon \pi
\end{eqnarray}
\begin{figure}
\centering
     \subfloat[\label{subfig-1:dummy}]{%
       \includegraphics[width=0.24\textwidth]{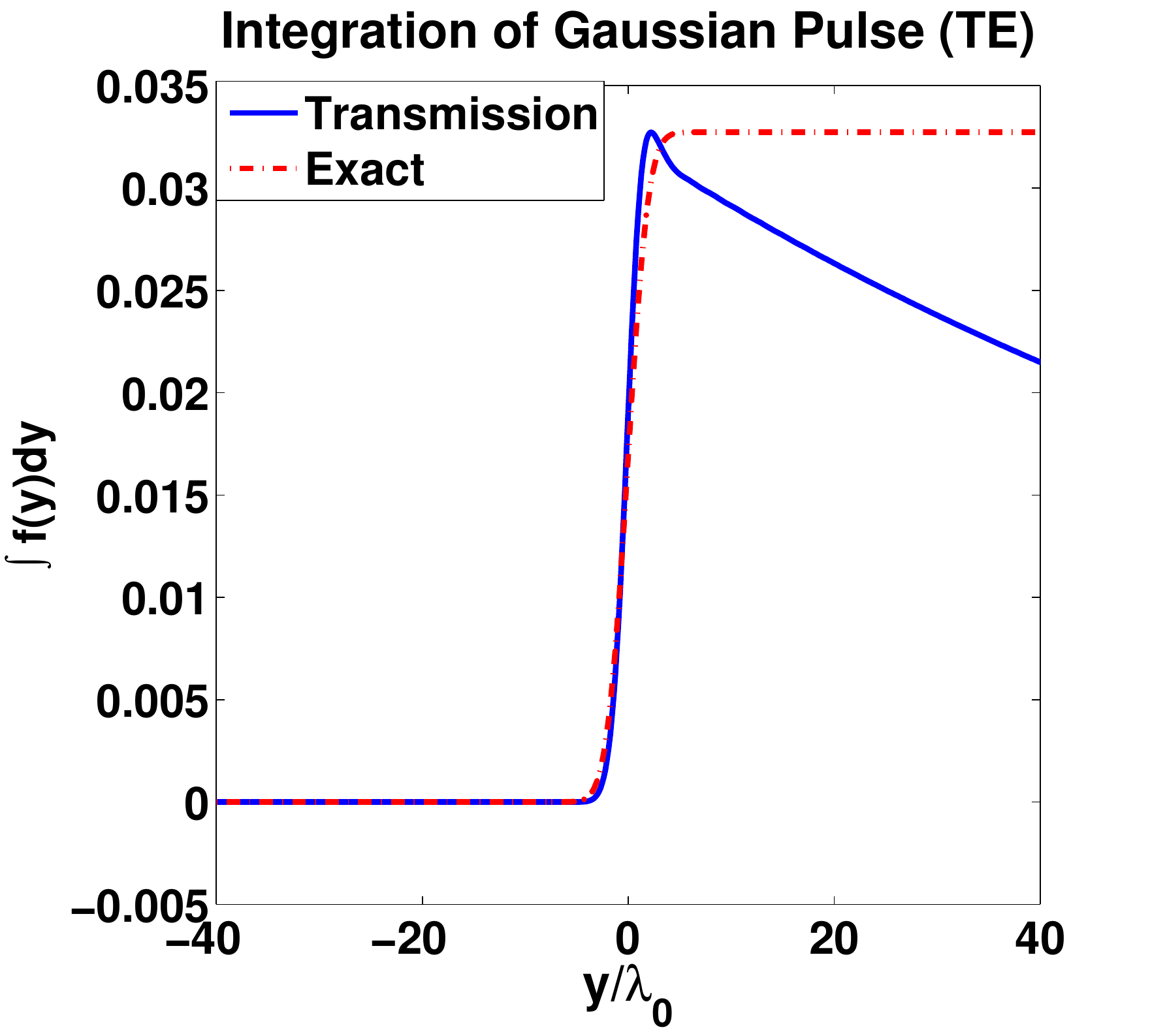}
     }
          \subfloat[\label{subfig-1:dummy}]{%
       \includegraphics[width=0.24\textwidth]{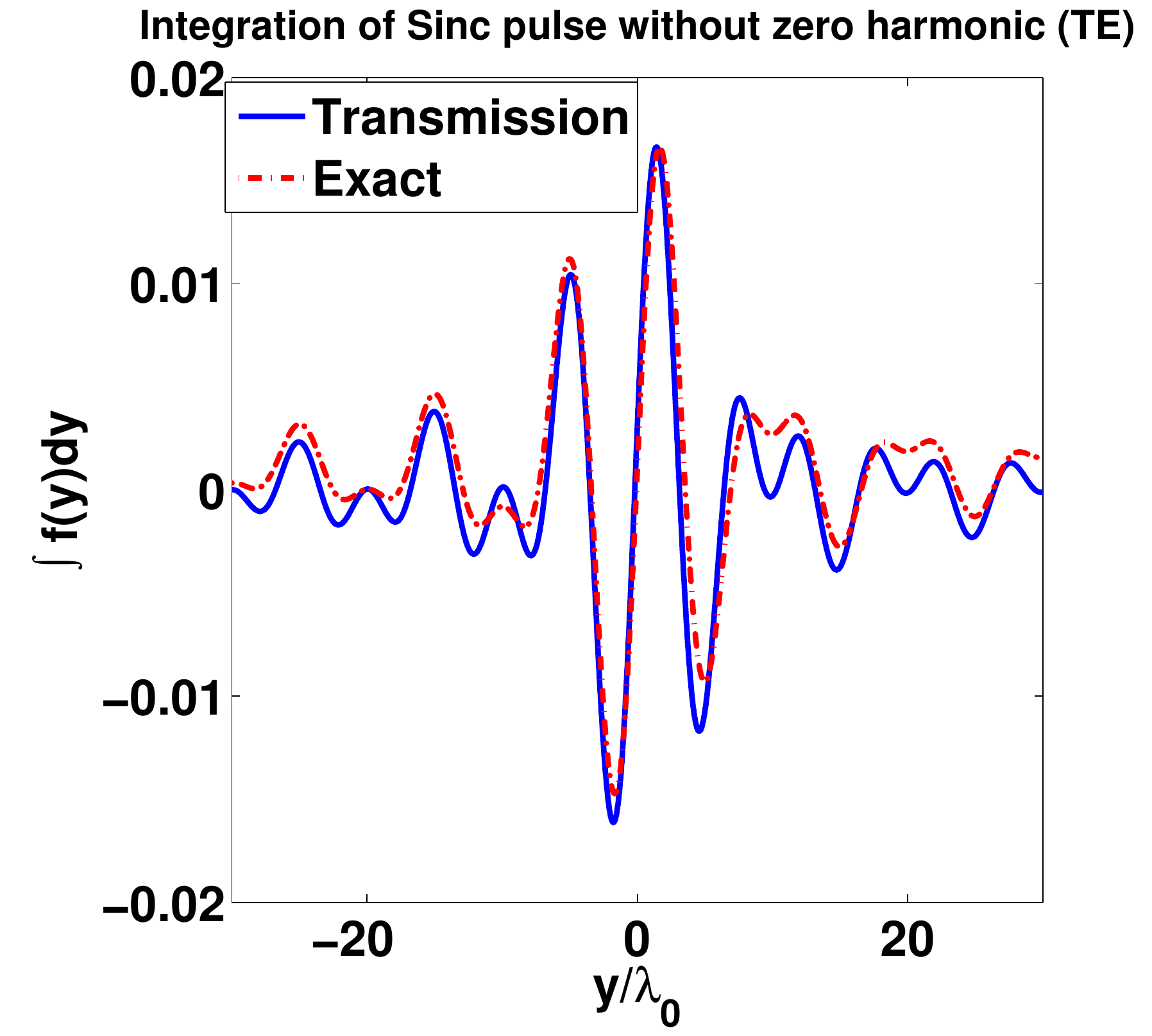}
     }
     \hfill
     \subfloat[\label{subfig-2:dummy}]{%
       \includegraphics[width=0.24\textwidth]{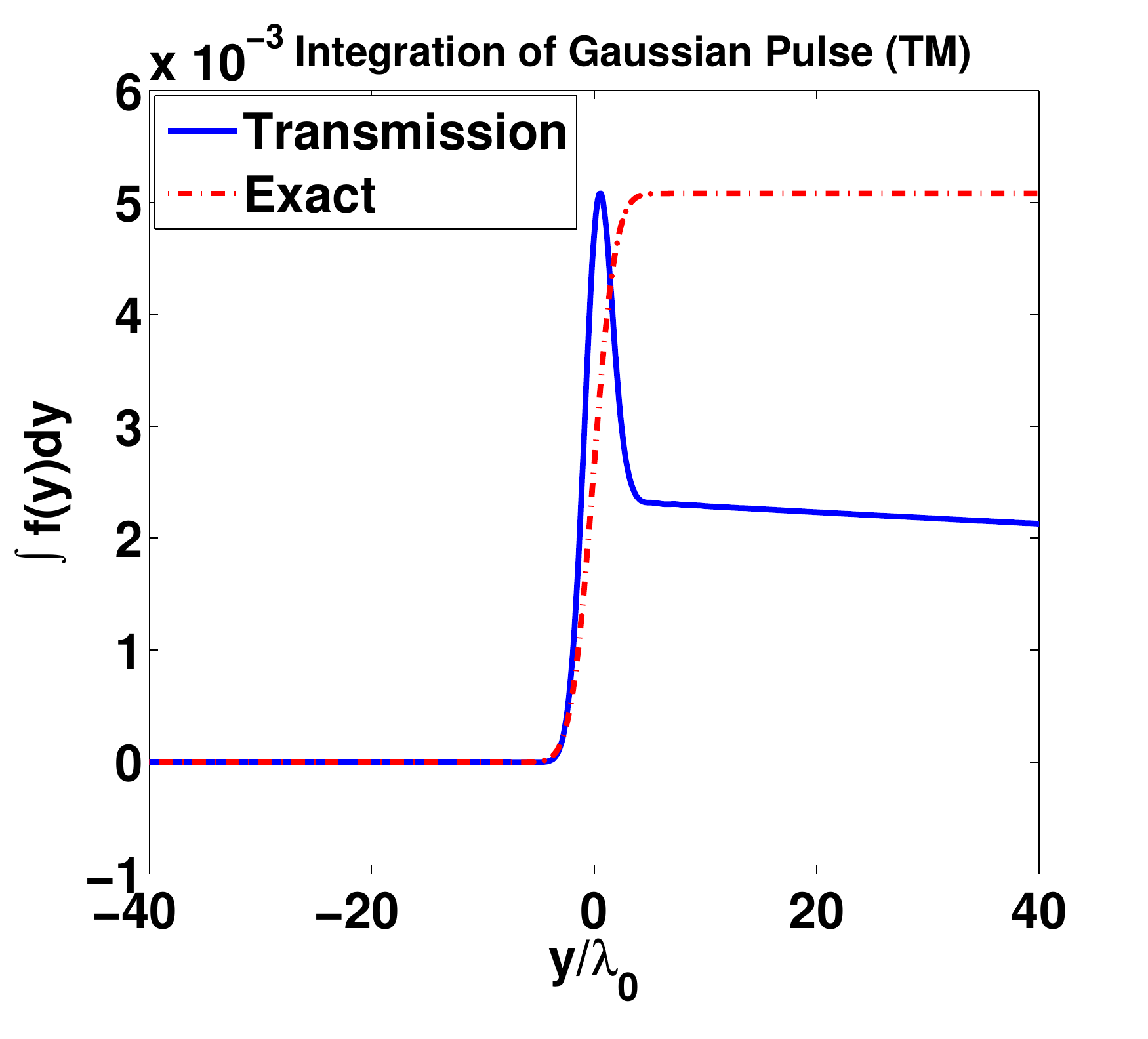}
     }
     \subfloat[\label{subfig-2:dummy}]{%
       \includegraphics[width=0.24\textwidth]{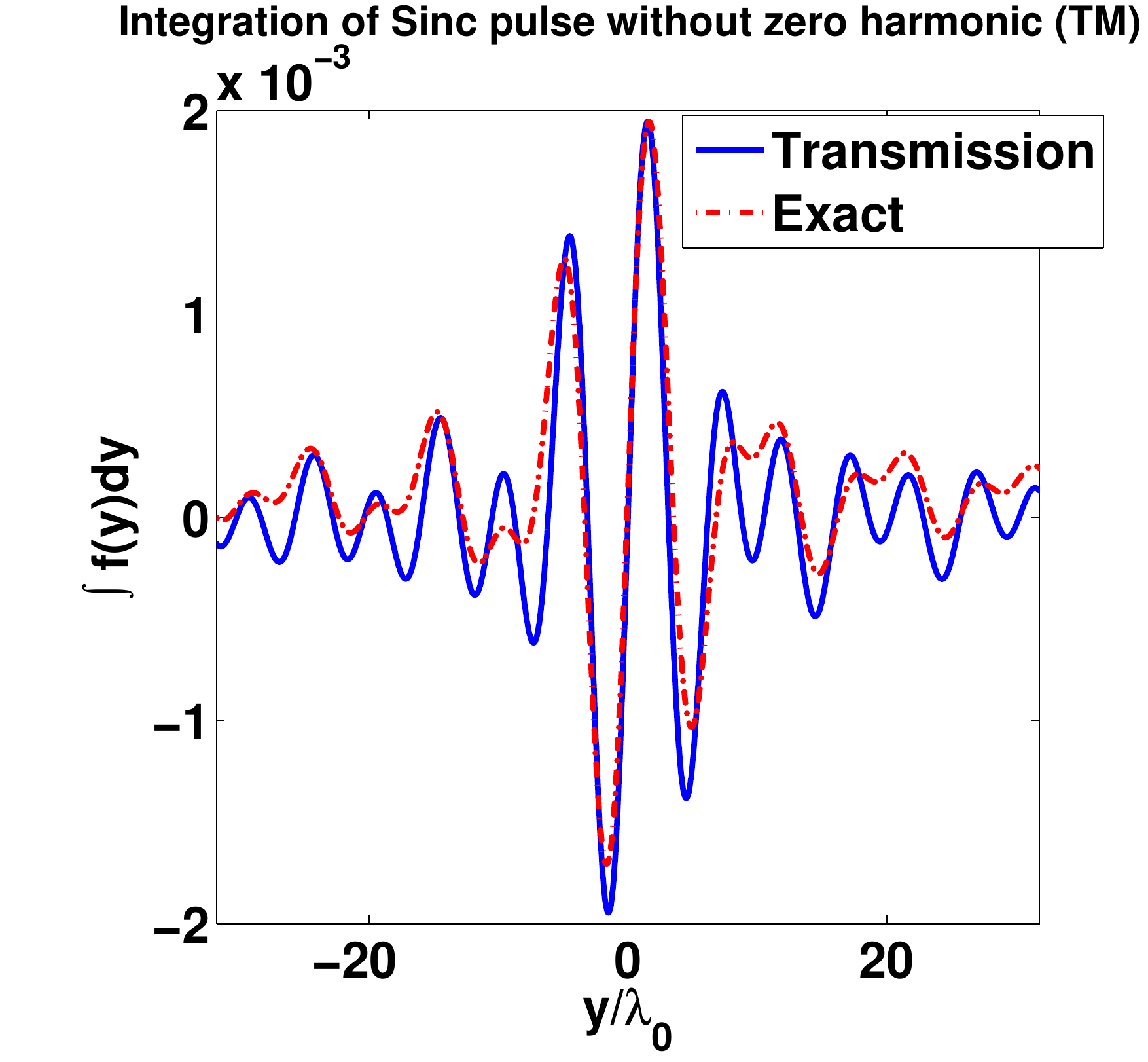}
     }
     \caption{Transmitted field (solid) together with scaled exact integration of the input filed (dash) for a Gaussian Pulse and Sinc profile without zero spatial harmonic: (a), (b) TE case (c),(d) TM case}
     \label{fig:dummy}
\end{figure}
\begin{figure}[!h]
\centering
     \subfloat[\label{subfig-1:dummy}]{%
       \includegraphics[width=0.3\textwidth]{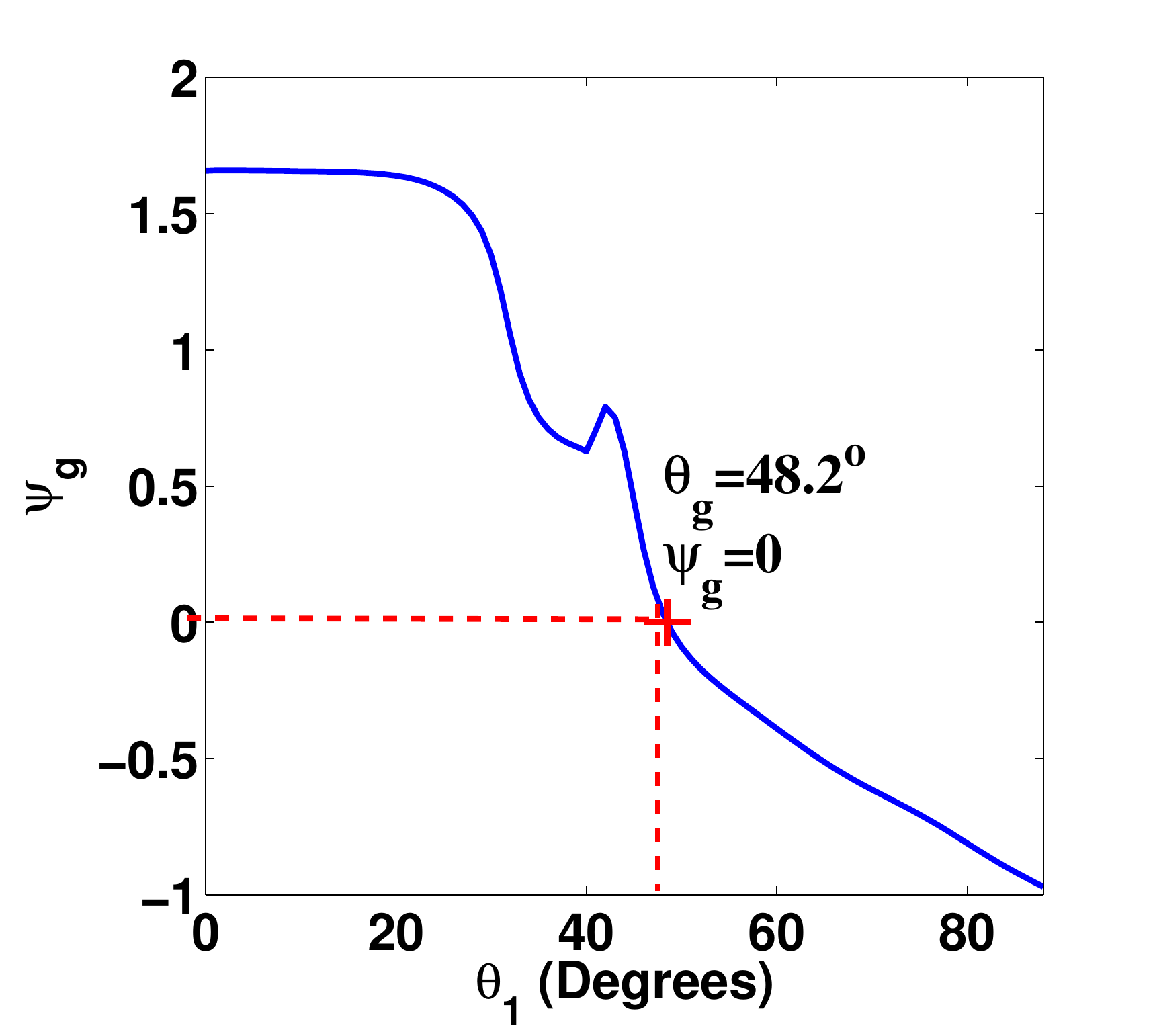}
     }
          \subfloat[\label{subfig-1:dummy}]{%
       \includegraphics[width=0.3\textwidth]{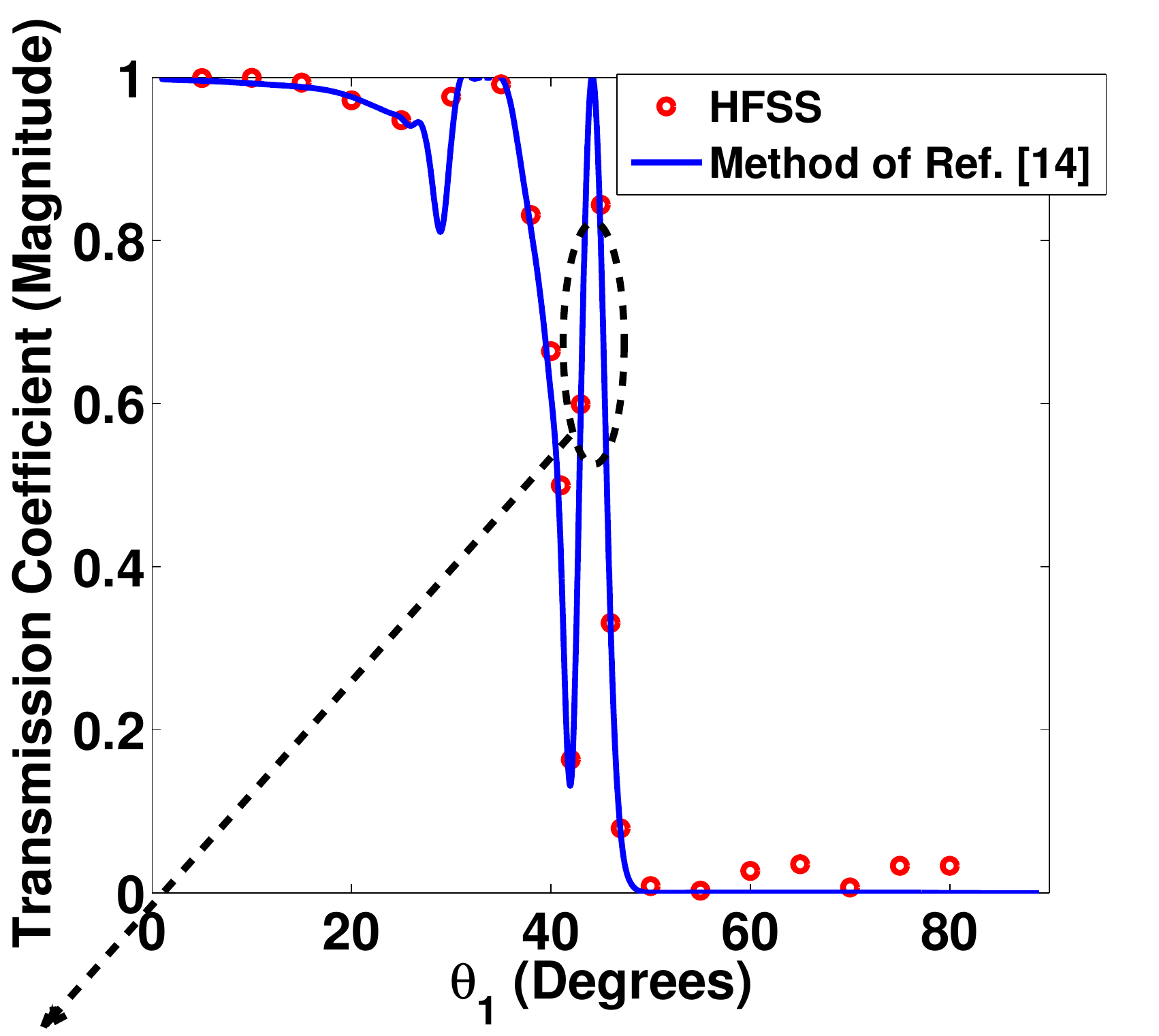}
     }
     \hfill
     \subfloat[\label{subfig-2:dummy}]{%
       \includegraphics[width=0.3\textwidth]{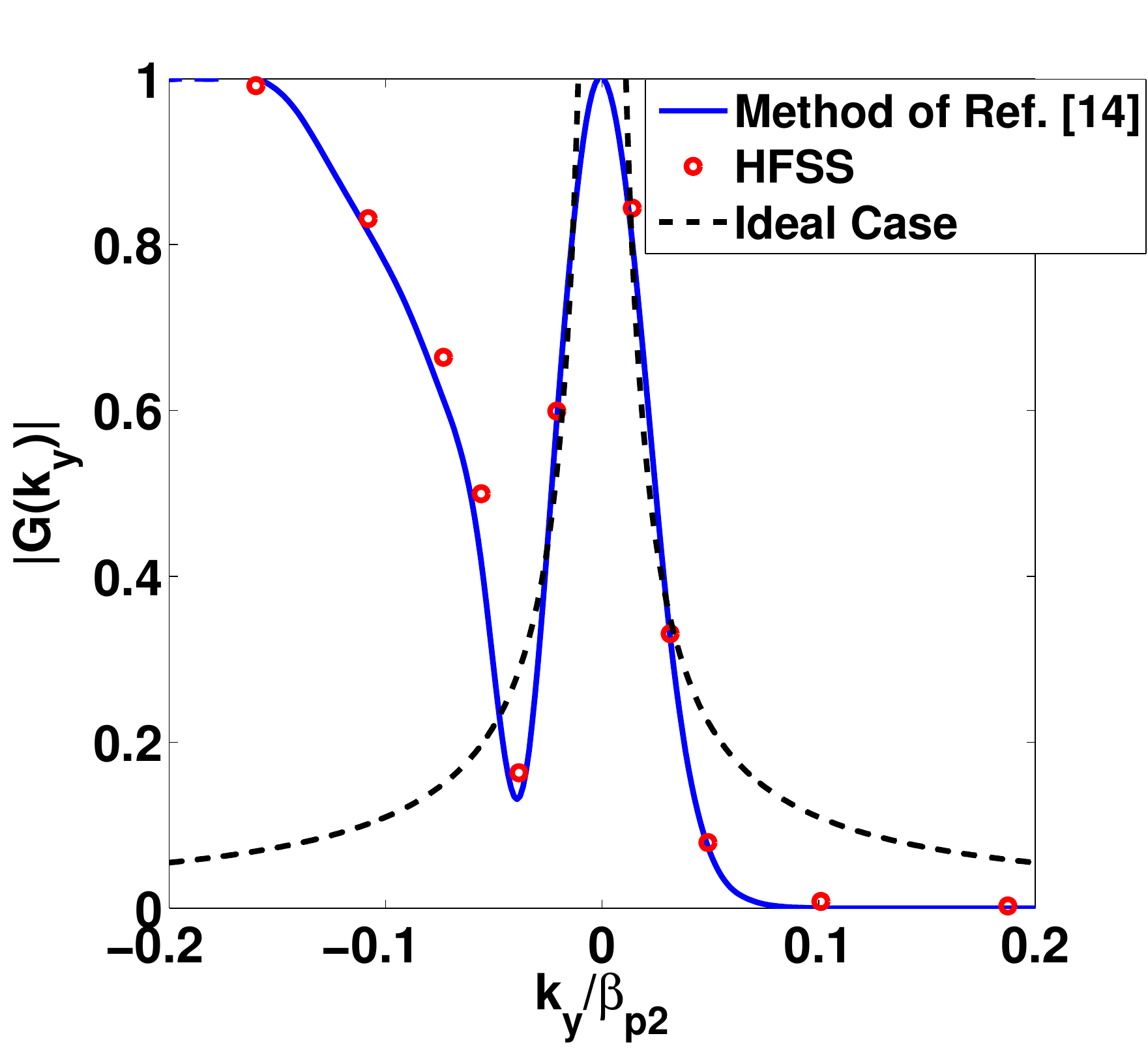}
     }
     \subfloat[\label{subfig-2:dummy}]{%
       \includegraphics[width=0.3\textwidth]{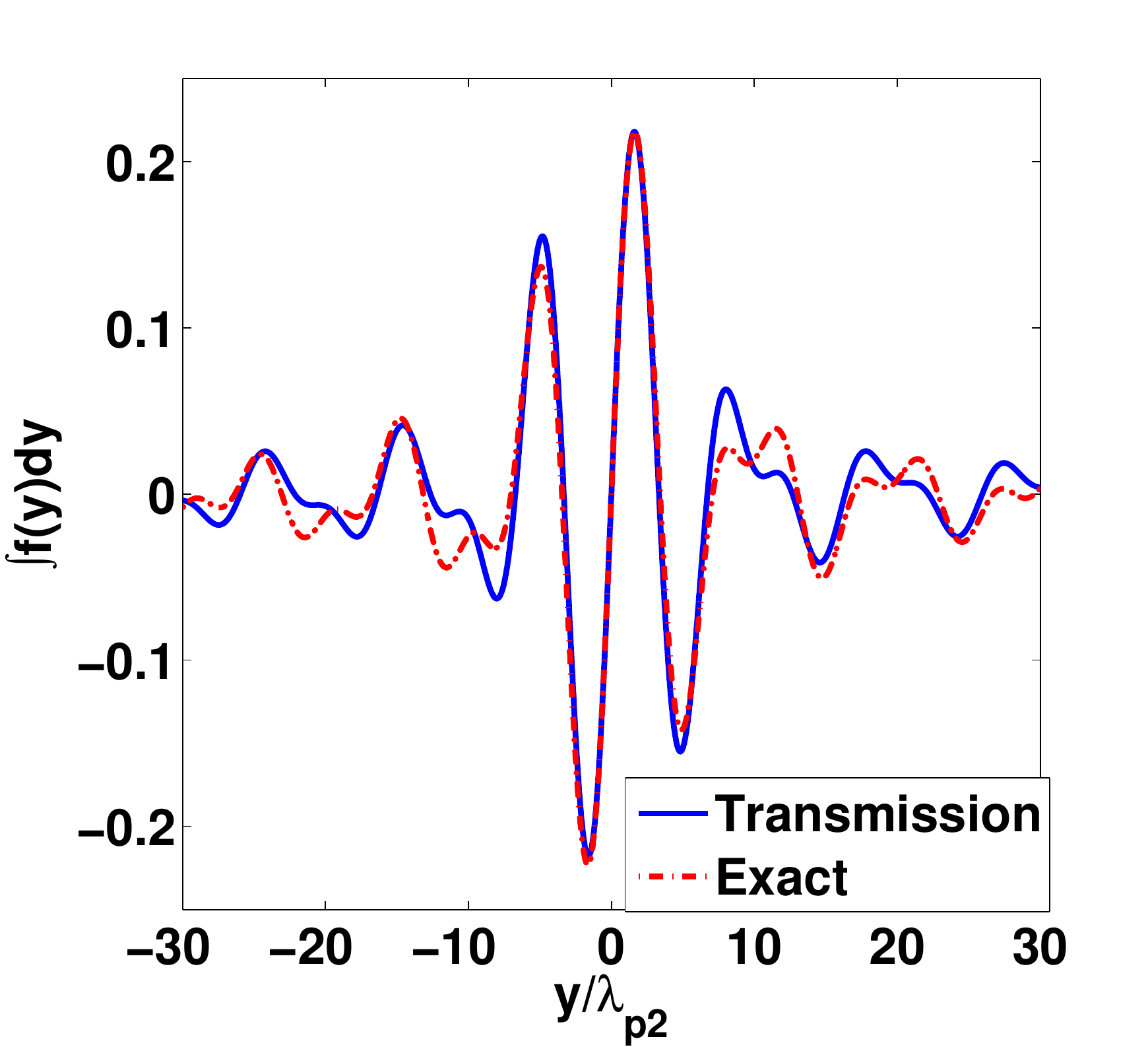}
     }
     \caption{Obtained results of the designed graphene-based integrator for $\beta_{p_{2}}=1.5~ \mu m^{-1}$: (a) Parameter $\psi_g$ versus the incident angle $\theta_1$, (b) Transmission coefficient of the integrator, (c) Green's Function of the integrator (d) Transmitted field together with scaled exact integration input filed }
     \label{fig:dummy}
\end{figure}
\begin{figure}[!h]
\centering
     \subfloat[\label{subfig-1:dummy}]{%
       \includegraphics[width=0.3\textwidth]{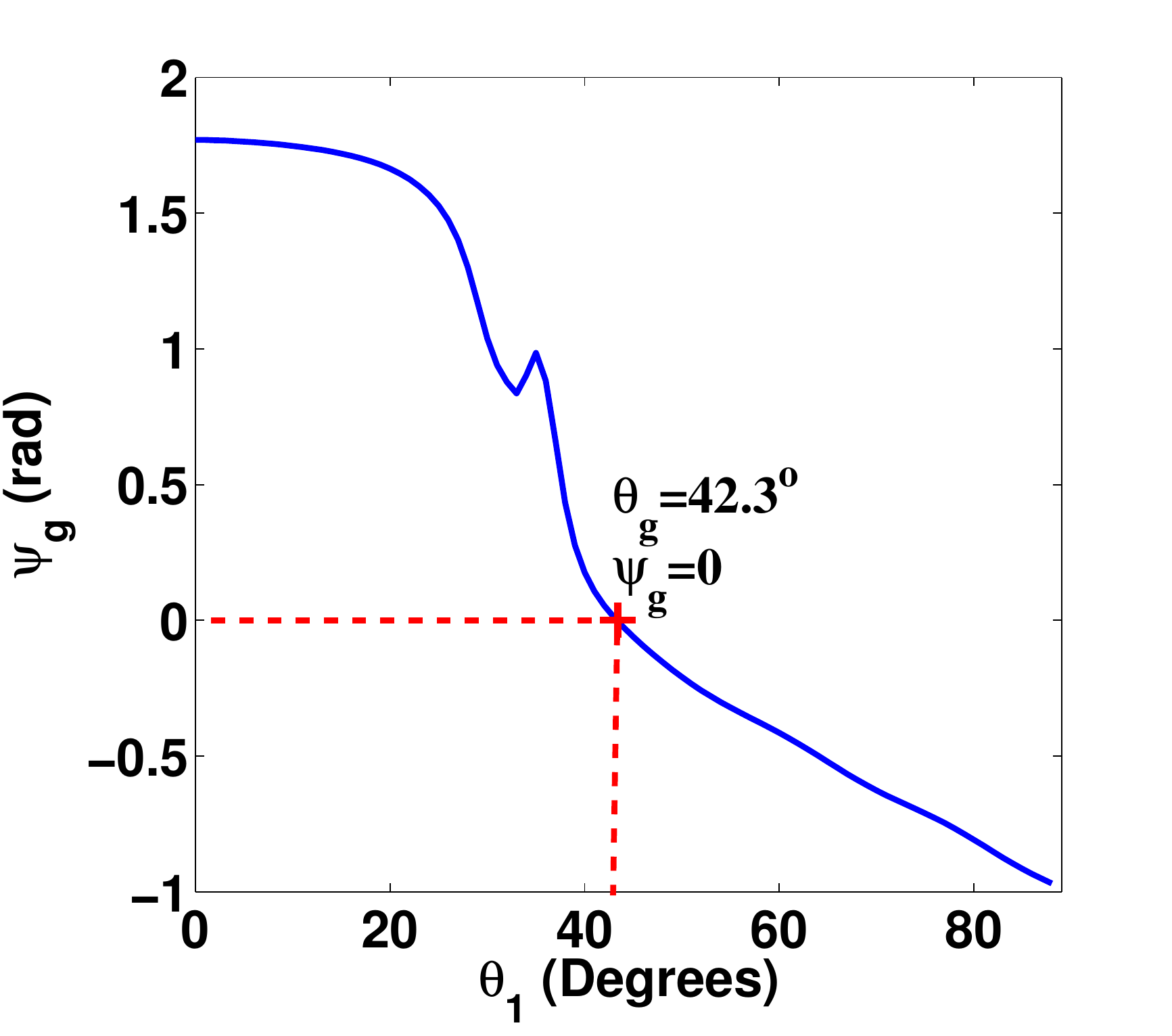}
     }
          \subfloat[\label{subfig-1:dummy}]{%
       \includegraphics[width=0.3\textwidth]{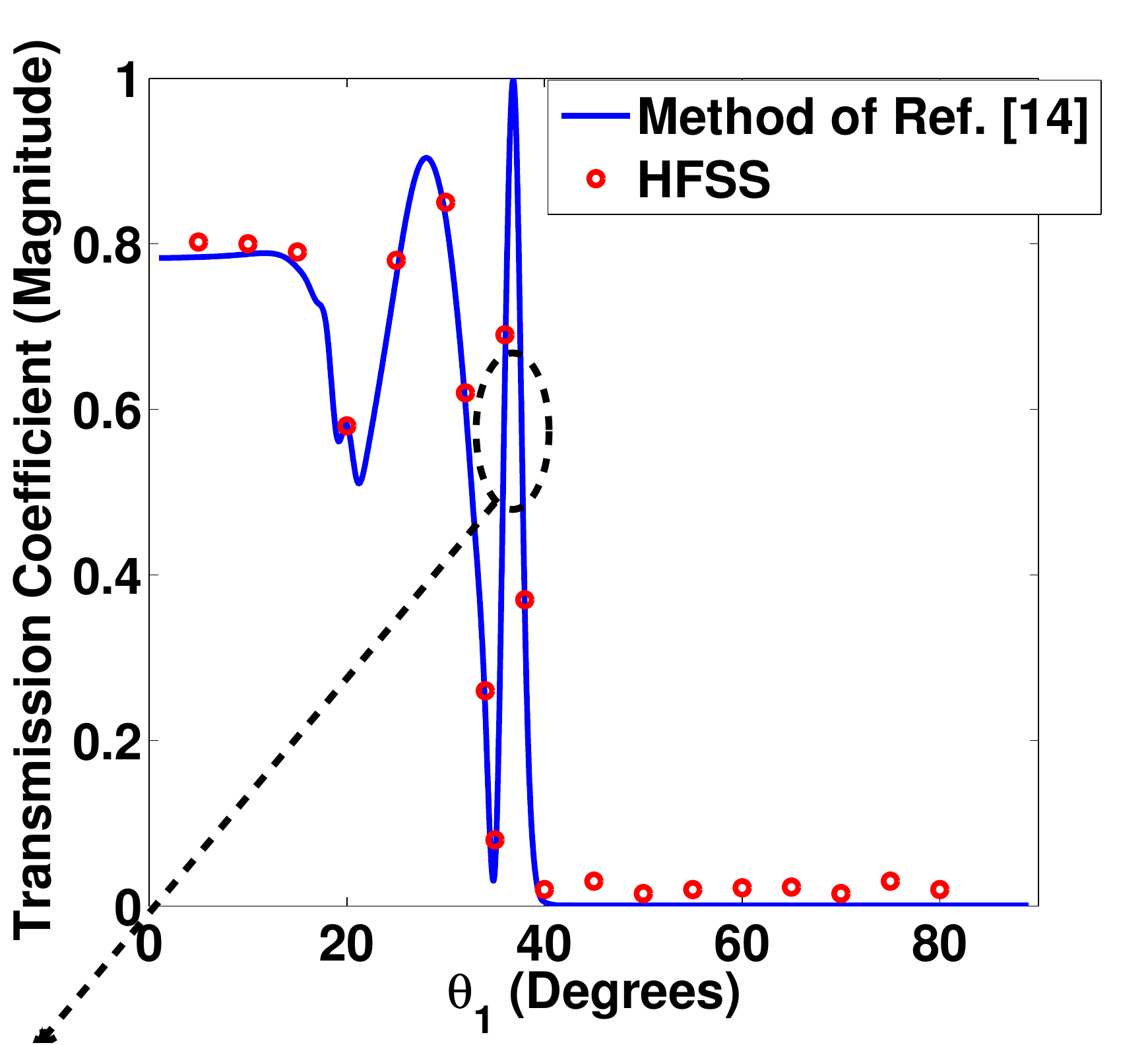}
     }
     \hfill
     \subfloat[\label{subfig-2:dummy}]{%
       \includegraphics[width=0.3\textwidth]{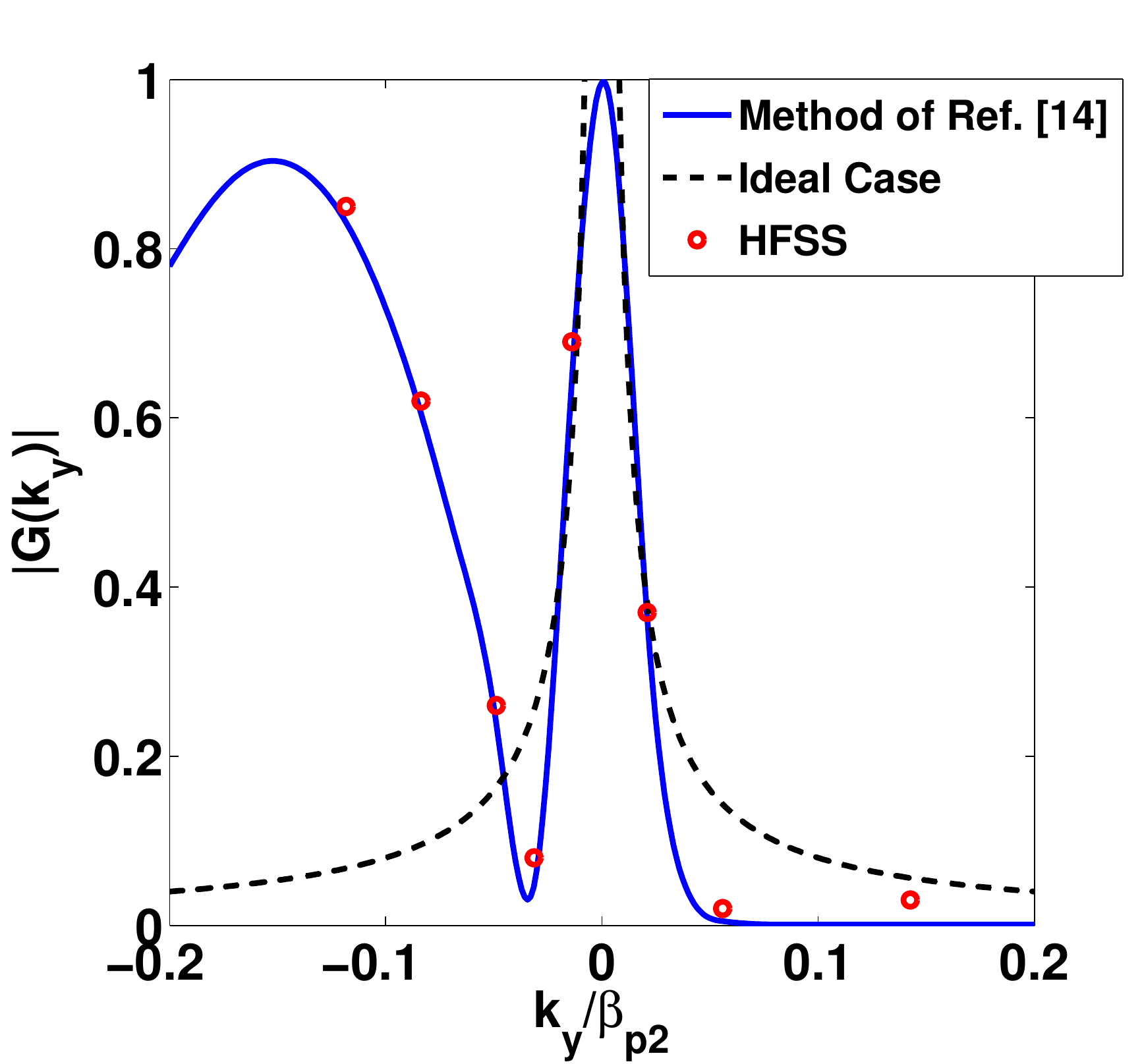}
     }
     \subfloat[\label{subfig-2:dummy}]{%
       \includegraphics[width=0.3\textwidth]{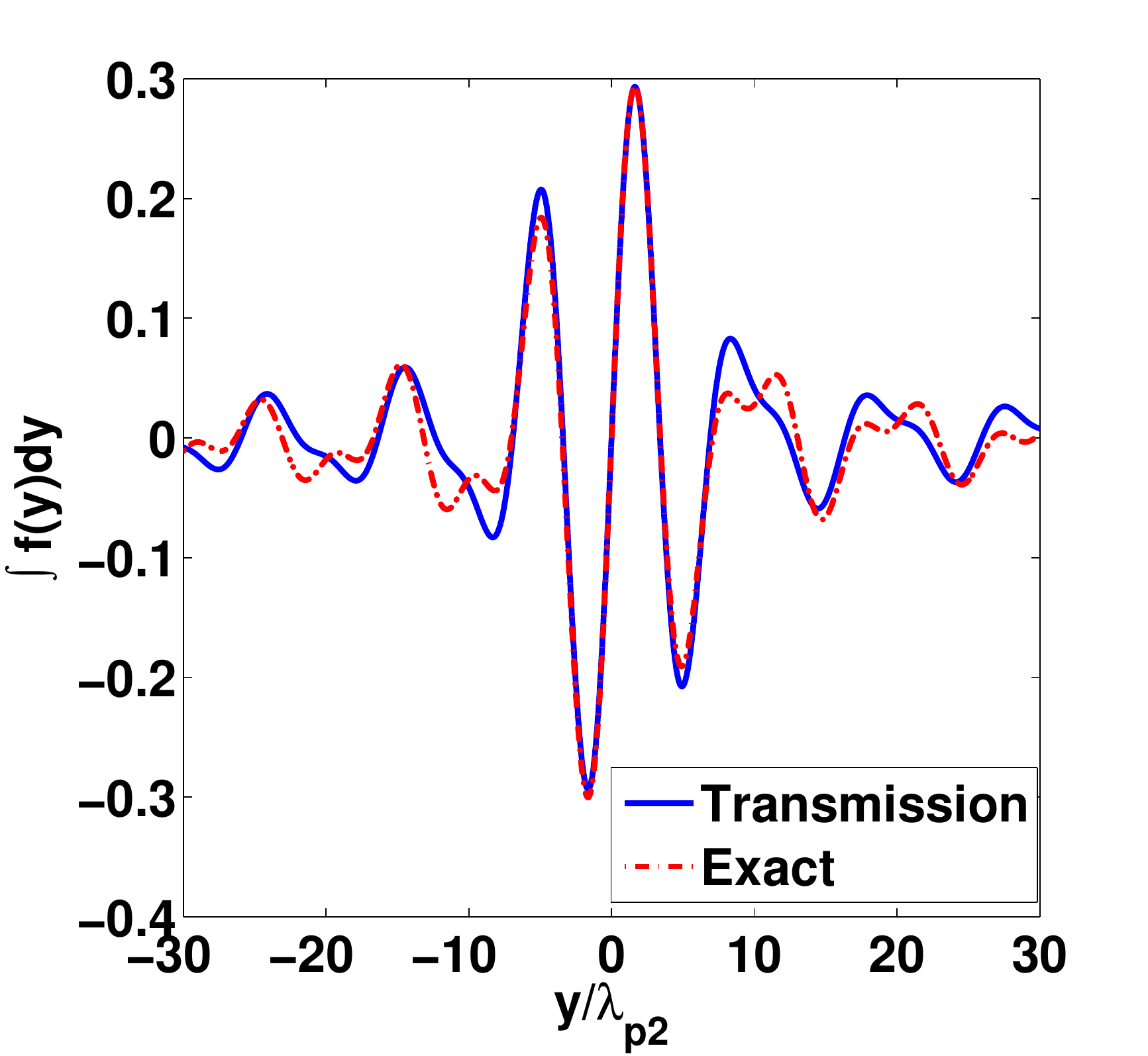}
     }
     \caption{Obtained results of the designed graphene-based integrator for $\beta_{p_{2}}=1.75~ \mu m^{-1}$: (a) Parameter $\psi_g$ versus the incident angle $\theta_1$, (b) Transmission coefficient of the integrator, (c) Green's Function of the integrator (d) Transmitted field together with scaled exact integration input filed}
     \label{fig:dummy}
\end{figure}
where $\varphi_{c}$ is the reflection phase shift on total reflection from the boundary between graphene strips with surface conductivities of $\sigma_{g_{1}}$ and $\sigma_{g_{2}}$, and $\upsilon$ is an integer.\\
\indent Fig. 4 (a) depicts the parameter $\psi_g$ versus the incident angle $\theta_1$. It can be observed that  the criteria of Eq. 4 has been established at the mode excitation angle of $\theta_{g}=48.2^{\circ}$.\\
 \indent The transmission coefficient of the integrator is calculated and depicted in Fig. 4 (b). The results are obtained by solving the integral equation in \cite{13} governing the behavior of surface charge density, and by HFSS. As it can be seen in this figure, the corresponding transmission coefficient has a first order pole at the incident angle of $\theta_1=44.8^{\circ}$, which is slightly lower than our design target ($\theta_{g}=48.2^{\circ}$). The slight difference is because of the fact that the criteria of Eq. 4 which has mainly been obtained using a ray-optic approach \cite{11}, is not accurate enough to model the propagation of graphene plasmons exactly \cite{15}. Therefore, to make the structure perform integration, the angle of incidence must be chosen to be $\theta_1=44.8^{\circ}$. The corresponding Green's function $G(k_y)$ is shown Fig. 4 (c). As expected, $G(k_y)$ has a pole at $k_y=0$. Therefore, like the previous case, our structure will be able to perform integration. \\
\indent An incident field with Sinc profile and spatial bandwidth of $W=0.1\beta_{p_2}$ whose its zero harmonic is eliminated, is then
considered as the input field.  The transmitted field along with the scaled exact integration are plotted in Fig. 4 (d). As expected, the results are in excellent agreement.\\
\indent To assess the reconfigurability of the integrator, we then assume that $\sigma_{2}$ is tuned such that $\beta_{p_{2}}=1.75~ \mu m^{-1}$ \cite{13}, and repeat similar procedures to the previous case. The parameter $\psi_g$, the associated transmission coefficient, the corresponding Green's function $G(k_y)$ and the transmitted field $t(y)$, are shown in Figs. 5 (a), (b), (c), (d), respectively. As it is observed, by changing the graphene surface conductivity, the incident angle for which our structure performs integration has been adapted to $\theta_1=37.7^{\circ}$. This validates the reconfigurability of the integrator. It should be noted that with respect to other spatial integrators like those proposed in \cite{4,9}, our proposed integrator is not only tunable but also highly miniaturized as the lengthes $h$ and $d$ are about 30 times smaller than the vacuum wavelength, and also there is no need to the two previously mentioned additional sub-blocks in metasurface approach.\\
\indent In summary, we proposed a new approach to realize spatial integrators using mode excitation. We first realized the integration employing a dielectric slab waveguide and exciting its corresponding mode. Inspired by this integrator, and making use of the unique features of graphene, we then demonstrated a reconfigurable and highly miniaturized integrator.

\bibliography{sample}

    \end{document}